\documentstyle[prl, aps, twocolumn, graphics, epsfig,
floats]{revtex}

\begin{document}

\title{A feasible ``Kochen-Specker'' experiment with single particles.
}

\author{Christoph Simon$^1$, Marek Zukowski$^2$, Harald Weinfurter$^{3,4}$, and Anton Zeilinger$^1$}

\address {$^1$Institut f\"ur Experimentalphysik, Universit\"at Wien,
Boltzmanngasse 5, A-1090 Wien,  Austria\\ $^2$ Instytut Fizyki
Teoretycznej i Astrofizyki Uniwersytet Gda$\acute{n}$ski,
PL-80-952 Gda$\acute{n}$sk, Poland\\ $^3$  Sektion Physik,
Ludwig-Maximilians-Universit\"at, D-80797 M\"unchen, Germany\\
$^4$ Max-Planck Institut f\"ur Quantenoptik, D-85748 Garching,
Germany}

\date{\today}
\maketitle
\begin{abstract}
We present a simple experimental scheme which can be used to
demonstrate an all-or-nothing type contradiction between
non-contextual hidden variables and quantum mechanics. The scheme,
which is inspired by recent ideas by Cabello and
Garc\'{\i}a-Alcaine, shows that even for a single particle, path
and spin information cannot be predetermined in a non-contextual
way.

\end{abstract}


\vskip 1pc

\narrowtext

Most predictions of quantum mechanics are of a statistical nature,
predictions for individual events are probabilistic. The question as to whether one can go beyond quantum mechanics
in this respect, i.e. whether there could be hidden variables
determining the results of all individual measurements, has been
answered to the negative for {\it local} hidden variables by
Bell's theorem \cite{bell}. Locality means that in such theories
the results of measurements in a certain space-time region are
independent of what happens in a space-time region that is
spacelike separated, in particular independent of the settings of
a distant measuring apparatus.

Bell's theorem refers to a situation where there are two particles
and where the predictions of quantum mechanics are statistical.
Furthermore, even definite (non-statistical) predictions of
quantum mechanics are in conflict with a local realistic picture
for systems of three particles or more \cite{ghz}.

The Kochen-Specker (KS) theorem \cite{KS} states that {\it
non-contextual} theories (NCT) are incompatible with quantum
mechanics. Non-contextuality means that the value for an
observable predicted by such a theory does not depend on the
experimental context, i.e. which other co-measurable observables
are measured simultaneously. In quantum mechanics, observables
have to commute in order to be co-measurable. Non-contextuality is
a more stringent demand than locality because it requires mutual
independence of the results for commuting observables even if
there is no spacelike separation.

So far there has not been an experimental test of
non-contextuality based on the original formulation of the KS
theorem, which refers to a single spin-1 particle (cf.
\cite{mermin}). However, experimental tests of local hidden
variable theories, such as tests of Bell's inequality and of the
GHZ paradox \cite{ghz}, can also be seen as tests of NCT. Note that such
experiments in general involve several particles.

Recently, in a very interesting paper Cabello and
Garc\'{\i}a-Alcaine (CG) \cite{cab} have proposed an experimental
test of the KS theorem based on two two-level systems (qubits).

In this paper we present a simple experimental scheme to test
non-contextuality which is inspired by the CG argument. The
experiment can be realized with single particles, using both their path
and their spin degrees of freedom. It leads to a non-statistical test of
non-contextuality versus quantum mechanics. In this respect it is
similar to the GHZ argument against local realism.

In the following, we first show how a very direct experimental test of
non-contextuality can be found, then we discuss our operational
realization.

Consider four binary observables $Z_1, X_1, Z_2$, and $X_2$. Let
us denote the two possible results for each observable by $\pm 1$.
In a NCT these observables have predetermined non-contextual
values $+1$ or $-1$ for individual systems, denoted as $v(Z_1),
v(Z_2), v(X_1)$, and $v(X_2)$. This means e.g. that for an
individual system the result of a measurement of $Z_1$ will always
be $v(Z_1)$ irrespective of which other co-measurable observables
are measured with it. We will show that the existence of such
non-contextual values is incompatible with quantum mechanics.

Imagine an ensemble E of systems for which one always finds equal
results for $Z_1$ and $Z_2$, and also for $X_1$ and $X_2$.
(Clearly, in order for this statement to be meaningful, $Z_1$ and
$Z_2$, and $X_1$ and $X_2$ have to be co-measurable.) In a NCT
this means that
\begin{equation}
 v(Z_1)=v(Z_2)\hspace{5mm}\mbox{and}\hspace{5mm} v(X_1)=v(X_2)
 \label{ensemble}
\end{equation}
 for each individual system of the
ensemble. Then there are only two possibilities: either
$v(Z_1)=v(X_2)$, which implies $v(X_1)=v(Z_2)$; or $v(Z_1)\neq
v(X_2)$, which implies $v(X_1) \neq v(Z_2)$. We will see that this
elementary logical deduction is already sufficient to establish a
contradiction between NCT theories and quantum mechanics.

To this end, let us express the above argument in a slightly
different way. Eq. (\ref{ensemble}) can be written as
\begin{equation}
v(Z_1)v(Z_2)=v(X_1)v(X_2)=1.
\end{equation}
Multiplying by $v(X_2)v(Z_2)$ it immediately follows that
\begin{equation}
v(Z_1)v(X_2)=v(X_1)v(Z_2). \label{pred1}
\end{equation}
Let us now introduce the notion of product observables such as
$Z_1X_2$. By definition, one way of measuring $Z_1X_2$ is to
measure $Z_1$ and $X_2$ separately and multiply the results; in
general, there are other ways. In particular, if another
compatible observable (e.g. $X_1Z_2$, cf. below) is measured
simultaneously, it will in general not be possible to obtain
separate values for $Z_1$ and $X_2$. However, in a non-contextual
theory, the result of a measurement of an observable must not
depend on which other observables are measured simultaneously.
Therefore the predetermined value $v(Z_1X_2)$, for example, in a
NCT has to follow the rule \cite{cab}
\begin{equation}
v(Z_1X_2)=v(Z_1)v(X_2).
\end{equation}
In this new language, our above argumentation can be resumed in
the following way:
\begin{equation}
 v(Z_1Z_2)=v(X_1X_2)=1\Rightarrow v(Z_1X_2)=v(X_1Z_2)
\label{pred}
\end{equation}
i.e. if our systems have the property expressed in Eq.
(\ref{ensemble}), then the two product observables $Z_1X_2$ and
$X_1Z_2$ must always be equal in a NCT. Note that in
general this prediction of NCT can only be tested if $Z_1X_2$ and
$X_1Z_2$ are co-measurable.

It follows from the results of \cite{cab} that the prediction
(\ref{pred}) leads to an observable contradiction with quantum
mechanics. To see this, consider a system of two qubits and the
observables \cite{cab}
\begin{equation}
Z_1:=\sigma_z^{(1)}, X_1:=\sigma_x^{(1)},
 Z_2:=\sigma_z^{(2)}, X_2:=\sigma_x^{(2)},
\end{equation}
  where
$\sigma_z^{(1)}$ means the z-component of the ``spin'' of the
first qubit etc. It is easy to check that this set of observables
satisfies all the properties required above. In particular, while
$Z_1$ and $X_1$, and $Z_2$ and $X_2$, do not commute, the two
product observables $Z_1X_2$ and $X_1Z_2$ do. Furthermore, the
quantum-mechanical two-qubit state
\begin{eqnarray}
|\psi_1\rangle&=&\frac{1}{\sqrt{2}}(|+z\rangle|+z\rangle+|-z\rangle|-z\rangle)\nonumber\\
&=&\frac{1}{\sqrt{2}}(|+x\rangle|+x\rangle+|-x\rangle|-x\rangle)
\label{state}
\end{eqnarray}
 is a joint eigenstate of the
commuting product observables $Z_1Z_2$ and $X_1X_2$ with both
eigenvalues equal to $+1$. Therefore, on the one hand the ensemble
described by this state possesses the property of the ensemble $E$
discussed above (cf. (\ref{ensemble})): the measured values of
$Z_1Z_2$ and $X_1X_2$ are equal to $+1$ for every individual
system. On the other hand, quantum mechanics predicts for the
state $|\psi_1\rangle$, that the measured value of $Z_1X_2$ will
always be opposite to the value of $X_1Z_2$. This can be seen by
decomposing $|\psi_1\rangle$ in the basis of the joint eigenstates
of the two commuting product observables $Z_1X_2$ and $X_1Z_2$:
\begin{equation}
|\psi_1\rangle=\frac{1}{\sqrt{2}}(|\chi_{1,-1}\rangle+|\chi_{-1,1}\rangle),
\label{sup}
\end{equation}
with

\begin{eqnarray}
|\chi_{1,-1}\rangle&=&\frac{1}{2}(|+z\rangle|+z\rangle+|-z\rangle|-z\rangle\nonumber\\
&&+|+z\rangle|-z\rangle-|-z\rangle|+z\rangle)\nonumber\\
&=&\frac{1}{\sqrt{2}}(|+z\rangle|+x\rangle-|-z\rangle|-x\rangle)\nonumber\\
&=&\frac{1}{\sqrt{2}}(|-x\rangle|+z\rangle+|+x\rangle|-z\rangle)
\end{eqnarray}
\begin{eqnarray}
|\chi_{-1,1}\rangle&=&\frac{1}{2}(|+z\rangle|+z\rangle+|-z\rangle|-z\rangle\nonumber\\
&&-|+z\rangle|-z\rangle+|-z\rangle|+z\rangle)\nonumber\\
&=&\frac{1}{\sqrt{2}}(|+z\rangle|-x\rangle+|-z\rangle|+x\rangle)\nonumber\\
&=&\frac{1}{\sqrt{2}}(|+x\rangle|+z\rangle-|-x\rangle|-z\rangle).
\end{eqnarray}

and
\begin{eqnarray}
Z_1X_2 |\chi_{1,-1}\rangle = +|\chi_{1,-1}\rangle\nonumber\\
X_1Z_2 |\chi_{1,-1}\rangle
=
-|\chi_{1,-1}\rangle\nonumber\\ Z_1X_2 |\chi_{-1,1}\rangle =
-|\chi_{-1,1}\rangle\nonumber\\ X_1Z_2 |\chi_{-1,1}\rangle =
+|\chi_{-1,1}\rangle \label{eigenv}
\end{eqnarray}
From (\ref{sup}) and (\ref{eigenv}) one sees that $|\psi_1\rangle$
is a linear combination of exactly those joint eigenstates of
$Z_1X_2$ and $X_1Z_2$ for which the respective eigenvalues are
opposite, which means, of course, that in a joint measurement the
two observables will always be found to be different. With Eq.
(\ref{pred}) in mind, this implies that the ensemble described by
$|\psi_1\rangle$ cannot be described by any non-contextual theory.

Note that one would already have a contradiction if quantum
mechanics only predicted that the observed values of $Z_1X_2$ and
$X_1Z_2$ are sometimes different, but in fact the result is even
stronger, with QM and NCT predicting exactly opposite results.
Thus, we have conflicting predictions for observable effects on a
non-statistical level \cite{nonstat} (cf. \cite{ghz}).

According to the argument presented in the previous paragraph, an
experimental test of non-contextuality can be performed in the
following way: (i) Show that $Z_1Z_2=1$ and $X_1X_2=1$ for systems
prepared in a certain way. (ii) Determine whether $Z_1X_2$ and
$X_1Z_2$ are equal for such systems. Note that in steps (i) and
(ii) the observables $Z_1, X_1, Z_2,$ and $X_2$ appear in two
different contexts.

Quantum mechanics predicts that step (i) can be accomplished by
constructing a source of systems described by the state
$|\psi_1\rangle$ and measuring $Z_1Z_2$ and $X_1X_2$ on these
systems. According to QM, both $Z_1Z_2$ and $X_1X_2$ will always
be found to be equal to $+1$. This can e.g. be verified by
measuring the pairs $Z_1$ and $Z_2$ and $X_1$ and $X_2$ separately
on many systems, and obtaining the values of $Z_1Z_2$ and $X_1X_2$
by multiplication. Alternatively, one could also perform joint
measurements of $Z_1Z_2$ and $X_1X_2$ on individual systems, but
for step (i) such joint measurements are not strictly necessary.
On the other hand, step (ii) definitely requires a joint
measurement of $Z_1X_2$ and $X_1Z_2$, because both negative and
positive values are to be expected for $Z_1X_2$ and $X_1Z_2$, and
we have to determine whether their values are equal (as required
by NCT) or opposite (as predicted by QM) for individual systems.

One might argue that the above alone is sufficient to demonstrate
in a very direct way the contextuality of quantum mechanics and
that, in view of the numerous experiments confirming the
predictions of quantum mechanics concerning entangled states
\cite{azrmp} of the kind of Eq. (\ref{state}), an experiment may
not be necessary. Nevertheless, an explicit operational
realization could have instructive advantages. Therefore we now
discuss a simple and intuitive experiment which would be readily
realizable.

\begin{figure}
        \includegraphics[width= 0.7 \columnwidth] {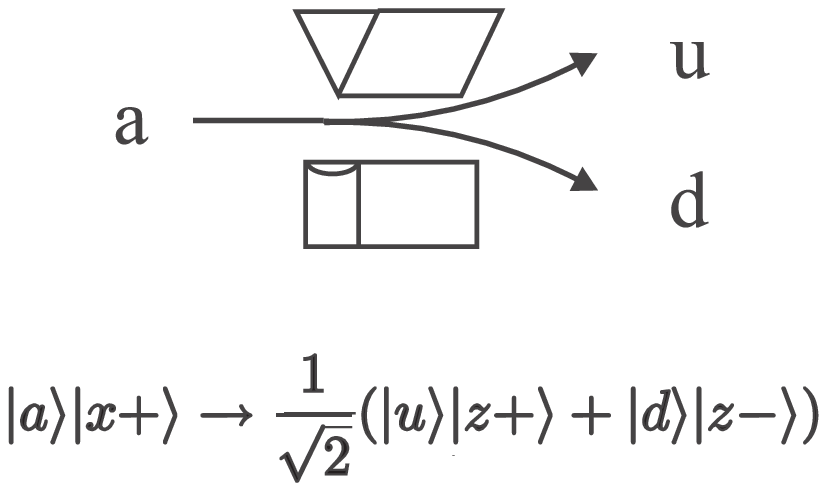}
        \caption{ Possible way of creating the single-particle version of $|\psi_1\rangle$
        given in Eq. (\ref{psi1}) using a standard Stern-Gerlach apparatus. A single particle with
        spin state $|x+\rangle=\frac{1}{\sqrt{2}}(|z+\rangle+|z-\rangle)$, i.e. spin
        along the positive $x$ direction, comes in from the left (spatial mode $|a\rangle$).
        By the Stern-Gerlach device, which separates incoming states according to the $z$-components
        of their spin, this is transformed into the desired superposition state.
        The outputs $u$ and $d$ could be
connected
        to the inputs of the devices of Figures \ref{separate} or \ref{joint}}
\label{source}
\end{figure}

One could consider realizing the above protocol with two entangled
particles, each one representing one of the qubits. Yet when
considering contextuality, non-locality is not an issue. This is
underlined by the observation that our contradiction arises for
joint measurements of the two qubits. Therefore the two necessary
qubits can also be carried by the same single particle.

In our scheme, the first qubit is emulated by the spatial modes of
propagation (paths) of a single spin-1/2 particle or photon, and
the second qubit by its spin (or polarization) degree of freedom
\cite{marek}. Spin-1/2 and photon polarization are completely
equivalent for our purposes. Our setup requires a source of
polarized single particles, beam splitters, and Stern-Gerlach type
devices. In practice, the experiment would be easiest to do with
photons because all these elements are readily available, in
particular polarized single-photon states can be produced to
excellent approximation via parametric down-conversion
\cite{singlephot}. Nevertheless, we will use the spin language in
the sequel because it is more familiar to most physicists.

Consider a situation where the particle can propagate in two
spatial modes $u$ and $d$, and let $|z+\rangle, |z-\rangle$ etc.
denote the particle's spin states as before. Then the state
$|\psi_1\rangle$ of Eq. (\ref{state}) is mapped onto the
one-particle state
\begin{equation}
\frac{1}{\sqrt{2}}(|u\rangle|z+\rangle+|d\rangle|z-\rangle).
\label{psi1}
\end{equation}
In Fig. \ref{source} we show how such a state can be prepared
experimentally.

\begin{figure}
\includegraphics[width=\columnwidth] {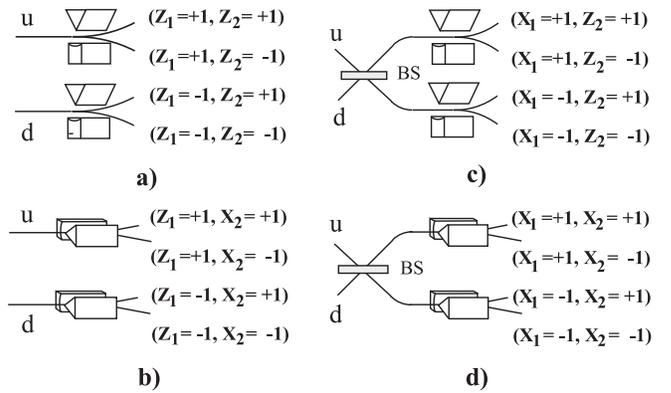}
\caption{  Devices for measuring pairs of the single-particle
observables of Eq. (\ref{obs}). A particle comes in from the left.
Note that in general the incoming states will have components in
both spatial modes $u$ and $d$ and of different spin. The devices
shown measure: a) $Z_1$ and $Z_2$; b) $Z_1$ and $X_2$; c) $X_1$
and $Z_2$; d) $X_1$ and $X_2$. BS in c) and d) stands for a
$50-50$ beam-splitter (see main text), which changes the basis of
path analysis from $|u\rangle, |d\rangle$, corresponding to a
measurement of $Z_1$, to $|u'\rangle, |d'\rangle$, thus leading to
a measurement of $X_1$. In a) and c) the Stern-Gerlach apparatus
are oriented along the $z$-axis (measurement of $Z_2$), in b) and
d) along the $x$-axis (measurement of $X_2$).}

\label{separate}
\end{figure}

The observables $Z_1,X_1,Z_2,X_2$ are now represented by
\begin{eqnarray}
Z_1&=&|u\rangle\langle u|-|d\rangle \langle d|\nonumber\\
X_1&=&|u'\rangle\langle u'|-|d'\rangle \langle d'|\nonumber\\
Z_2&=&|z+\rangle\langle z+|-|z-\rangle \langle z-| \nonumber\\
X_2&=&|x+\rangle\langle x+|-|x-\rangle\langle x-|, \label{obs}
\end{eqnarray}
where $|u'\rangle=\frac{1}{\sqrt{2}}(|u\rangle+|d\rangle),
|d'\rangle=\frac{1}{\sqrt{2}}(|u\rangle-|d\rangle),|x+\rangle=
\frac{1}{\sqrt{2}}(|z+\rangle+|z-\rangle),
|x-\rangle=\frac{1}{\sqrt{2}}(|z+\rangle-|z-\rangle)$, i.e. $u'$
and $d'$ denote the output modes of a 50-50 beam-splitter with
inputs $u$ and $d$, and $|x+\rangle$ and $|x-\rangle$ are the spin
eigenstates along the $x$ direction . Clearly, $Z_1$ and $X_1$ act
on the path, and $Z_2$ and $X_2$ on the spin degree of freedom.

\begin{figure}
        \includegraphics[width= \columnwidth] {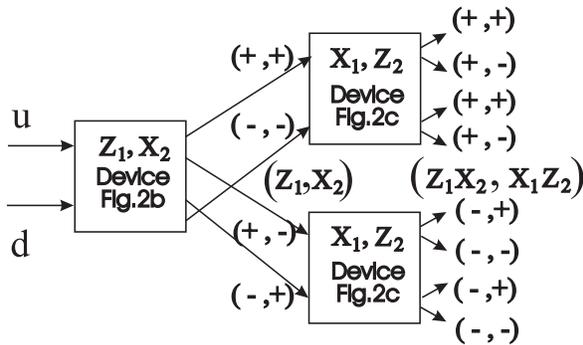}
        \caption{Device for performing a joint measurement of $Z_1X_2$ and
        $X_1Z_2$.
         A device
        performing a joint measurement of $Z_1Z_2$ and $X_1X_2$ can be constructed in an analogous way.
        Instead of leading to detectors, the outputs of the device of Fig.
\ref{separate}b, which measures $Z_1$ and $X_2$ (the connecting
beams are labeled by the signs of the eigenvalues of the two
observables), are now fed into two replicas of the device of Fig.
\ref{separate}c, which measure $X_1$ and $Z_2$. The output beams
are labeled by the signs of the eigenvalues of  $Z_1X_2$ and
$X_1Z_2$.  The first device separates the two eigenspaces of the
degenerate product observable $Z_1X_2$. Eigenstates of $Z_1X_2$
with eigenvalue +1 are sent up, those with eigenvalue -1 are sent
down.
 Detection of the particle behind one of the
two subsequent devices is a  measurement of $X_1Z_2$ and at the
same time completes the measurement of $Z_1X_2$.  For an ensemble
of particles with the property $Z_1Z_2=X_1X_2=1$ (which can be
verified using the devices of Fig. \ref{separate}a and
\ref{separate}d) quantum mechanics predicts that
 the
particles can emerge only via one of those four outputs for which
the values of $Z_1X_2$ and $X_1Z_2$ are {\it opposite}, i.e. the
second, fourth, fifth, and seventh from the top, whereas
non-contextual theories predict exactly the complementary set of
outputs.} \label{joint}
\end{figure}

In Fig. \ref{separate} we show the devices that measure pairs of
one-particle observables, such as $Z_1$ and $Z_2$. While any
device that performs a state analysis in the basis of common
eigenstates of $Z_1X_2$ and $X_1Z_2$ can be considered to perform
a joint measurement of these two observables,   the particular
realization presented in Fig. \ref{joint} has the merit of showing
explicitly that a joint measurement of two product observables is
performed. It also shows how the information that is obtained in
the first stage of the measurement about the values of $Z_1$ and
$X_2$ is partially erased such that only information about the
product $Z_1 X_2$ is retained, in order to make the measurement of
$X_1 Z_2$ in the second stage possible. Using the devices of Figs.
1-3 one can realize the full experimental procedure described
above.

The present scheme allows the simplest non-statistical
experimental test of non-contextuality that is known to us
\cite{zukwein}. Similarly to the original Kochen-Specker paradox
it requires only a single particle (though two degrees of
freedom). With the experimental setup consisting of a simple
interferometer, it shows particularly clearly that the appearance
of the paradox is related to the superposition principle.

We thank \v{C}. Brukner for very useful discussions. This work was
supported by the Polish-Austrian project 11/98b {\it Quantum
Communication and Quantum Information II (1998-1999)}, the
Austrian Science Foundation (FWF), projects S6504 and F1506, and
by the European Commission project {\it Long Distance Photonic
Quantum Communication} (IST-99-1-1A FET P1 QIPC). H.W. was also
supported by the APART program of the Austrian Academy of
Sciences.\\


\begin{references}

\bibitem{bell} J.S. Bell, Physics {\bf 1}, 195 (1965).
\bibitem{ghz} D.M. Greenberger, M. Horne, and A. Zeilinger, in
{\it Bell's Theorem, Quantum Theory, and Conceptions of the
Universe}, edited by M. Kafatos (Kluwer, Dordrecht, 1989); D.M.
Greenberger, M.A. Horne, A. Shimony, and A. Zeilinger, Am. J.
Phys. {\bf 58}, 1131 (1990)
\bibitem{KS} E.P. Specker, {\it Selecta}
(Birkh\"auser Verlag, Basel, 1990); S. Kochen and E.P. Specker, J.
Math. Mech. {\bf 17}, 59 (1967); J.S. Bell, Rev. Mod. Phys. {\bf
38}, 447 (1966);   A. Peres, J. Phys. A {\bf 24}, L175 (1991);
N.D. Mermin, Rev. Mod. Phys. {\bf 65}, 803 (1993). See also A.
Peres, {\it Quantum Theory: Concepts and Methods} (Kluwer Academic
Publishers, Dordrecht, The Netherlands, 1993).

\bibitem{mermin} C. Simon, \v{C}. Brukner, and A. Zeilinger, quant-ph/0006043;
see also N.D. Mermin, quant-ph/9912081

\bibitem{cab} A. Cabello and G. Garc\'{\i}a-Alcaine, Phys. Rev. Lett. {\bf 80},
1797 (1998). The CG version of the KS theorem also has the
distinguishing feature of being state independent, in the spirit
of the original Kochen-Specker theorem.



\bibitem{nonstat} Of course, in a real experiment visibilities are
never perfect, and one would have to use some kind of inequality
to rigorously establish the contradiction.

\bibitem{azrmp} A. Zeilinger, Rev. Mod. Phys. {\bf 71}, S288
(1999)

\bibitem{marek} M. Zukowski, Phys. Lett. A {\bf 157}, 198 (1991); M. Czachor, Phys. Rev. A {\bf 49}, 2231 (1994).

\bibitem{singlephot} S. Friberg, C.K. Hong, and L. Mandel, Phys. Rev. Lett. {\bf 54}, 2011 (1985)




\bibitem{zukwein} For a single-photon experiment that implements a statistical test
of NCT versus QM see M. Michler, H. Weinfurter, and M. Zukowski,
Phys. Rev. Lett. {\bf 84}, 5457 (2000).


\end{references}
\end{document}